\documentclass[a4paper,aps,pra,twocolumn,superscriptaddress,showpacs]{revtex4-1}
\usepackage{graphicx}
\usepackage{amsmath}
\usepackage{amsfonts}
\usepackage{color}
\usepackage{bm}
\usepackage{pdfcomment}
\usepackage{pgfplots}
\usepackage{subfigure}

\newcommand{\beq}{\begin{equation}}
\newcommand{\eeq}{\end{equation}}
\newcommand{\comment}[1]{}

\newcommand{\eq}[1]{Eq.~\eqref{#1}}

\newcommand{\seq}[1]{Sec.~\ref{#1}}

\newcommand{\be}{\begin{equation}}
\newcommand{\ee}{\end{equation}}
\newcommand{\bem}{\begin{multline}}

\usepackage{pgfplots}
\usetikzlibrary{plotmarks}

\graphicspath{{figures/}}

\begin{document}


\title{Optomechanical many-body cooling using frustration}

\author{Thom\'as Fogarty$^*$}
\affiliation{Theoretische Physik, Universit\"at des Saarlandes, D-66123 
Saarbr\"ucken, Germany}
\affiliation{Quantum Systems Unit, OIST Graduate University, Onna, Okinawa 904-0495, Japan}

\author{Haggai Landa}
\affiliation{LPTMS, CNRS, Univ.~Paris-Sud, Universit\`e Paris-Saclay, 91405 Orsay, France}

\author{Cecilia Cormick}
\affiliation{IFEG, CONICET and Universidad Nacional de C\'ordoba, Ciudad Universitaria,
X5016LAE, C\'ordoba, Argentina}

\author{Giovanna Morigi}
\affiliation{Theoretische Physik, Universit\"at des Saarlandes, D-66123 
Saarbr\"ucken, Germany}

\date{\today}

\begin{abstract}
We show that the vibrations of an ion Coulomb crystal can be cooled to the zero-point motion through the optomechanical coupling with a high-finesse cavity. Cooling results from the interplay between coherent scattering of cavity photons by the ions, which dynamically modifies the vibrational spectrum, and cavity losses, that dissipate motional energy. The cooling mechanism we propose requires that the length scales of the crystal and the cavity are mismatched so that the system is intrinsically frustrated, leading to the formation of defects (kinks). When the pump is strong enough, the anti-Stokes sidebands of all vibrational modes can be simultaneously driven. These dynamics can be used to prepare ultracold chains of dozens of ions within tens of milliseconds in state-of-the-art experimental setups. In addition, we identify parameter regimes of the optomechanical interactions where individual localized modes can be selectively manipulated, and monitored through the light at the cavity output. These dynamics exemplify robust quantum reservoir engineering of strongly-correlated mesoscopic systems and could find applications in optical cooling of solids.

\end{abstract}

\pacs{03.67.Bg, 03.65.Yz, 42.50.Dv, 03.67.Mn}

\maketitle
\par
\section{Introduction} 

Optomechanical interactions enable the observation and control of quantum dynamics of massive objects, thereby opening perspectives for fundamental science and novel technologies \cite{Aspelmeyer:2014,Kippenberg:2008}. This regime is typically achieved by realizing high Q-factor mechanical resonators, which are well described as individual quantum harmonic oscillators \cite{Kippenberg:2008}. As such, they can be cooled to ultralow temperatures using standard techniques of laser spectroscopy like sideband cooling \cite{Wineland:1979,Stenholm:1986,Diedrich:1989,Eschner:2003}. Specifically, one uses the optomechanical coupling to resonantly drive the anti-Stokes (``red'') sideband of the mechanical resonator, thus transferring energy from the vibrational to the optical mode, with photon dissipation making the process irreversible. Cooling to the ground state is achieved when the corresponding cooling rate significantly exceeds the competing heating rates \cite{Wilson-Rae:2004,Wilson-Rae:2007,Marquardt:2007,Painter:2013}. 

Attempts to extend these procedures to optically cool all vibrational modes of a bulk material face a major problem in the  form of the coupling with the external environment \cite{Dunlop,Sheik}. As a result, optical cooling has so far only reached cryogenic temperatures in highly pure materials \cite{NatPhot:2010}. Moreover, even if the heating rates can be made sufficiently low, a straightforward extension of sideband cooling suffers from two fundamental caveats. In the first place, cooling the modes one by one leads to a cooling time that scales up with the number of modes. Furthermore, as the number of modes grows, spectrally resolving an individual mode becomes increasingly challenging \cite{Wineland:1998,Morigi:2001}, so that while a mode is being sideband-cooled the other modes are heated by the coupling to the same field. Significant improvement of the cooling of large structures thus requires one to develop novel concepts that go beyond a simple extension of the idea of controlling and cooling each individual component.  

In this paper we propose a route to successfully address these two problems, making use of the optomechanical coupling of a collection of scatterers to a high-finesse electromagnetic cavity. Our proposal requires that the individual scatterers are polarizable particles subjected to a periodic substrate potential -- the optical cavity mode. A key ingredient is to choose the periodic potential with a wavelength that is incommensurate with the interparticle distance. The length mismatch then leads to the formation of defects, called kinks, in the mean-field equilibrium configuration. Coherent scattering of cavity photons occurs at the kinks with a high rate and when the cavity field is strong enough to spectrally bunch the vibrational modes the entire system can be cooled to the ground state.
Our suggestion thus makes use of optomechanical cavity cooling developed in \cite{Vuletic:2000, Ritsch:2013, Aspelmeyer:2014} combined with the dynamical modification of the vibrational spectrum in the Frenkel-Kontorova model, which describes the essential features of static friction in one dimension \cite{FK,Braun}. 

 \begin{figure*}[hbt]
\begin{center}
\includegraphics[width=0.99\textwidth]{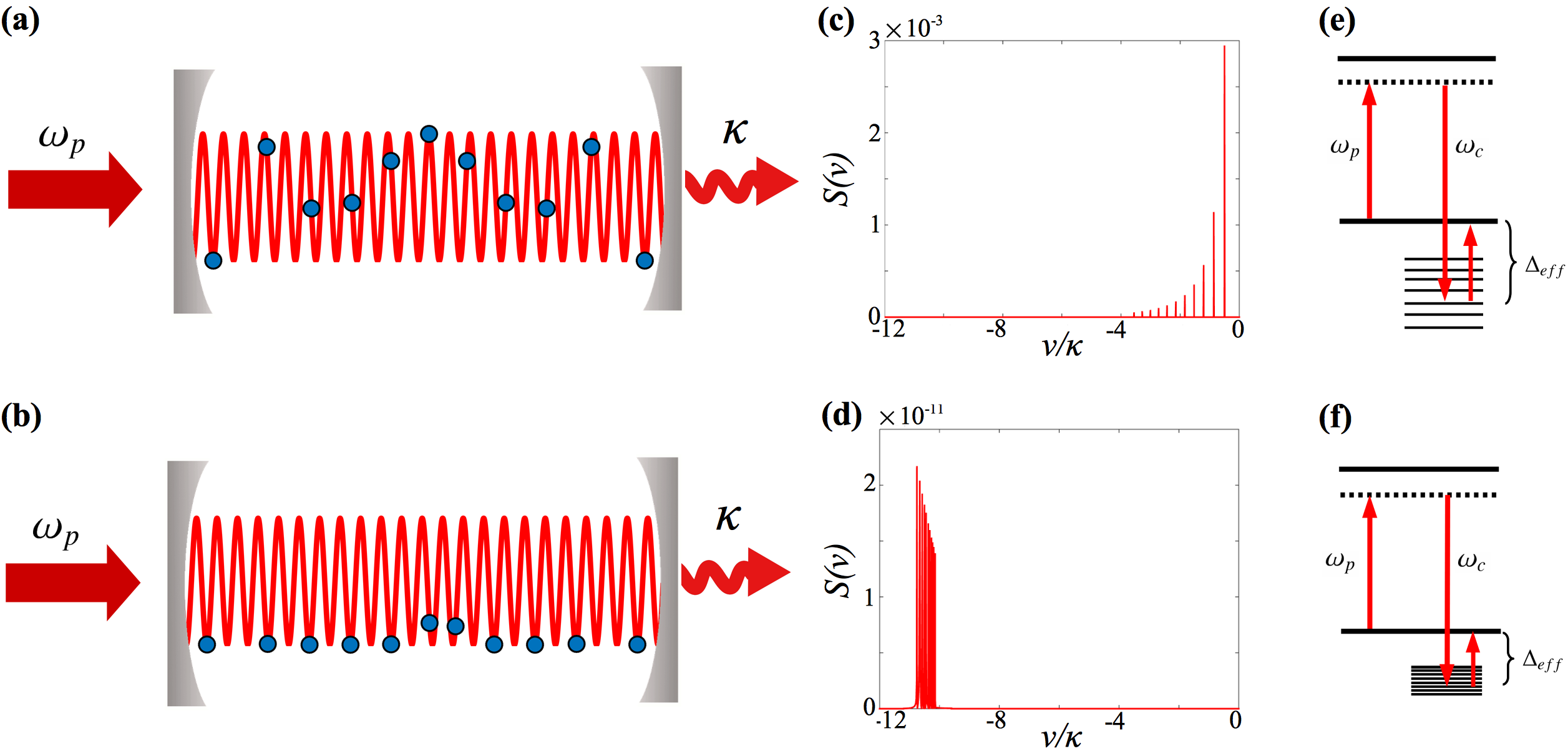}
\caption{
The axial vibrations of an ion chain are cooled to the ground state through the optomechanical coupling with a cavity. The cavity standing-wave mode at frequency $\omega_c$ is pumped by a laser at frequency $\omega_p<\omega_c$ and dissipates photons at rate $\kappa$. The figure shows a schematic representation of particle positions in (a) the sliding phase and (b) the pinned phase in the optical potential; these phases are found for incommensurate lengths of the ion chain and the cavity wavelength. Due to this competition, in the pinned phase defects called kinks are formed, which are essential for coupling all modes simultaneously to the cavity. (c) For low pump strengths (here $\eta=1.5\kappa$) the particles are in the sliding phase and the vibrational frequencies can be individually resolved, as visible in the spectrum at the cavity output, $S(\nu)$ (in units of $1/$Hz), as a function of the frequency shift $\nu$ from the pump frequency (see the Appendix for details and \seq{parameters} for the parameters). Each mode can thus be separately cooled by resonant coherent scattering processes when a pump photon at frequency $\omega_p$ is absorbed and a cavity photon at frequency $\omega_c$ is emitted, carrying away the energy of a vibrational excitation (determined by $\Delta_{\rm eff}$, see text for details), as sketched in (e). For a strong pump (here $\eta=300\kappa$) the particles are in the pinned phase, the bandwidth of the modes (d) much smaller than the cavity linewidth $\kappa$ and (f) coherent scattering can simultaneously drive all anti-Stokes sidebands, thus cooling the chain. 
 \label{schematic}}
\end{center}\end{figure*}

Fig.~\ref{schematic} illustrates our proposal, which we analyze in detail for a system of trapped ions as a candidate platform to test these dynamics. We consider an array of equally charged ions confined by an external trap and brought to crystallization by means of laser Doppler cooling \cite{Birkl:1992,Raizen:1992,Dubin:1999}. These systems, in fact, constitute a controllable form of condensed matter, where the binding forces result from the interplay of the external potential and the Coulomb repulsion, while the coupling with the external environment is minimized. Ground-state cooling by sequential sideband cooling of each axial vibrational mode was demonstrated for chains of up to four ions \cite{Sackett:2000,Eschner:2003,Roos:2008}, but the extension of these techniques to longer chains faces the challenges of achieving the required spectral resolution and overcoming the heating rates.
Following realizations of precise control in the coupling of trapped ions to an optical cavity mode \cite{Cetina:2013,Casabone:2012,Linnet:2012}, ion chains have proven extremely suitable for the exploration of the rich physics of the Frenkel-Kontorova model in different parameter regimes \cite{Garcia-Mata:2007,Haeffner:2011, Benassi:2011,Tosatti:2013, Mandelli:2013, Vuletic:2015}. 
We base our study on the recent analysis of the stationary states of an ion chain in a setup where the Frenkel-Kontorova model is generalized to include the dynamical degree of freedom of the cavity field \cite{Cormick:2012,Cormick:2013,Fogarty:2015}.
Using state-of-the-art experimental parameters, we show that it is possible to cool an ion chain of several tens of ions to the zero-point motion in time scales on the order of tens of milliseconds.

We note that selective methods for shaping the mode spectrum of an ion crystal have also been studied, employing anharmonic traps \cite{isospaced,isospaced2}, impurities \cite{SKaler}, Rydberg-excited states \cite{Li} and discrete solitons with localized modes \cite{solitons08,solitons10,landa} that have also been observed experimentally \cite{solitonsexpa,solitonsexpb, solitonsexpc,solitonsexpd,solitons13,solitons13b, solitons15}. The use of a cavity as discussed in the present work has the advantage of providing a tunable mechanism that combines mode shaping with tailored dissipation, thus providing an example of robust quantum reservoir engineering.
The technique we propose must also be considered within the perspective of existing laser cooling mechanisms for ion chains. Ground-state cooling of chains of few ions has been experimentally realized by sequentially sideband-cooling each axial vibrational mode \cite{Sackett:2000,Eschner:2003,Roos:2008}. Other ground-state cooling strategies have been proposed as an alternative to mode-by-mode sideband cooling with trapped ions. These rely on modifying the scattering cross-section of the individual scatterers, either by means of an appropriately tailored spatial gradient shifting the resonance frequency \cite{Wunderlich:2005} or by using quantum interference with a technique called EIT cooling \cite{Morigi:2000,Eschner:2003,Lin:2013, EIT-Roos}. Using this dynamics, recently the radial modes of a chain of 18 ions have been cooled to the ground state \cite{EIT-Roos}. Differing from these methods, the optomechanical coupling we propose here enables the cooling of ordered structures of polarizable particles independent of their internal structure.

This article is organized as follows. In \seq{Sec:Dynamics} we review the theoretical model at the basis of our study. Prospects and parameters for cooling an ion chain to the zero-point motion are discussed in 
 \seq{Sec:Cooling}, while in \seq{Sec:CoolKink} we study cavity cooling and spectroscopy of a single kink mode. Concluding comments are presented in \seq{Sec:TheEnd}, where we also discuss the potential of applying these concepts to cooling other many-body systems. In the Appendix we provide more details of our calculations.

\section{Optomechanics with an ion chain}\label{Sec:Dynamics}

The system we consider is an array of ions whose dipoles strongly interact with a single standing-wave mode of a high-finesse resonator, as depicted in Fig. \ref{schematic} (a)-(b). There are $N$ ions of charge $q$ and mass $m$ and their equilibrium positions and motion are confined to the $x$-axis. We restrict to a one-dimensional model, as this allows us to simply calculate the mode shaping of the array by means of the Frenkel-Kontorova theory \cite{FK, Braun, Mukamel}. This assumption simplifies the treatment but is not essential to the dynamics we discuss. It is experimentally realized by means of a deep transverse confinement, such that there is a gap between the axial and the transverse excitations \cite{Morigi:2004}. The equilibrium positions of the ions, about which they perform small vibrations, result from the interplay between their mutual repulsion, the axial confinement, and the mechanical forces associated with coherent scattering of cavity photons \cite{Fogarty:2015}. Coherent scattering processes are modulated by the spatial intensity distribution $\cos^2(kx)$, where $\cos(kx)$ is the cavity spatial mode function along $x$ and $k$ is the mode wave-number. The intensity gradient of the scattering amplitude gives rise to a mechanical force which tends to trap the ions at the lattice minima. The depth of the lattice is proportional to the number of photons inside the cavity, a dynamical variable which depends on the interplay between the external pump and the resonator losses, as well as on the ions positions within the cavity mode. This gives rise to a nonlinear dependence on the depth of the lattice that traps the ions, which in turn modifies the ions positions \cite{Larson:2008,Fogarty:2015,Asboth}. This optical potential is also subjected to photon losses, which are a key element of the dynamics for cooling the chain's vibrations. 

In the following subsection, \ref{Sec:Model}, we review the basic theoretical model for the evolution of our system, while in \seq{Sec:Mean-field} we focus on the equilibrium configurations of the ions interacting with the cavity mode. We then analyze the fluctuations of ions and cavity field about the equilibrium configuration in \seq{Sec:Opto}. This theoretical model is the basis of the cooling scheme of \seq{Sec:Cooling}. We note that the content of this Section was developed in \cite{Cormick:2012,Cormick:2013} and is included here for completeness.

\subsection{Master equation}
\label{Sec:Model}

The coupled dynamics of the cavity and the ions is described by a master equation for the density matrix $\varrho$ of the ions' external degrees of freedom and the cavity mode,
\begin{equation}
\label{M:eq}
\partial_t\varrho=\frac{1}{{\rm i}\hbar}[H,\rho]+\mathcal L[a]\varrho.
\end{equation}
The Hamiltonian $H$ governs the coherent dynamics and $\mathcal L$ is the Lindbladian describing the incoherent processes, which we assume to consist of cavity losses at rate $\kappa$
\begin{equation}
\mathcal L[a]\varrho=\kappa(2a\varrho a^\dagger -a^\dagger a\varrho-\varrho a^\dagger a )\,,
\end{equation}
where we denoted by $a$ and $a^\dagger$ the annihilation and creation operators of a cavity photon.
The Hamiltonian $H$ can be decomposed into three parts: 
\begin{equation}
\label{eq:Htot}
H=H_{\rm ions}+H_{\rm cav}+H_{\rm int}\,, 
\end{equation}
where $H_{\rm ions}$ is the Hamiltonian in absence of the cavity field, and it consists of the ions' kinetic energy, the trap potential energy and the Coulomb repulsion:
\begin{eqnarray}
&&H_{\rm ions}=\sum_{j=1}^N\frac{p_j^2}{2m}+V_{\rm ions}\,,\\
&&V_{\rm ions}=\sum_{j=1}^N\frac{1}{2}m\omega_t^2x_j^2 + \sum_{j=1}^N\sum_{k=j+1}^N \frac{q^2}{4\pi\epsilon_0}\frac{1}{|x_j-x_k|}\,.
\end{eqnarray}
Here, the variables $x_j$ and $p_j$ are the canonically-conjugated position and momentum of ion $j$, $\omega_t$ is the trap frequency, and $\epsilon_0$ the vacuum permittivity. The Hamiltonian $H_{\rm cav}$ is the energy of the cavity field in absence of the ions:
\begin{equation}
H_{\rm cav}=-\hbar \Delta_ca^\dagger a+{\rm i}\hbar \eta (a^\dagger -a)\,,
\end{equation}
with $\eta$ the strength of the laser driving the cavity, which has the dimension of a frequency, and $\Delta_c=\omega_p-\omega_c$ the detuning between the laser pump and the cavity mode. Finally, the optomechanical coupling reads:
\begin{equation}
H_{\rm int}=\hbar U_0 a^\dagger a\sum_{j=1}^N\cos^2(kx_j)\,.
\end{equation}
This Hamiltonian describes the mechanical potential exerted by the cavity field on the ions, and conversely the frequency shift of the cavity mode depending on the ions positions. This term scales with $U_0$, which has the dimension of a frequency and is the amplitude of coherent scattering of a cavity photon by an ion. Since we will make reference to atomic ions, this regime is reached when the detuning $\Delta_d=\omega_p-\omega_d$ of the pump field  from the ion dipolar transition is much larger than the other typical frequency scales of the problem. Then,  $U_0=g^2/\Delta_d$, and its sign is determined by the sign of the detuning.

\subsection{Mean-field treatment}
\label{Sec:Mean-field}

By means of Doppler cooling one can remove thermal energy from the ions' motion, so that at steady state the kinetic energy is orders of magnitude smaller than the ions' potential energy. In this limit the ions perform small vibrations about the equilibrium positions $\{\bar{x}_j\}$, determined by the balance between the mechanical forces acting on the particles.  
The equilibrium positions are found by solving the coupled equations for the expectation values $\bar \zeta(t)={\rm Tr}\{\hat \zeta\varrho(t)\}$, with $\zeta=x_j,p_j,a$, where the time evolution is performed by solving the master equation \eqref{M:eq}:
\begin{align}
 & \frac{\partial~}{\partial t} \bar a =  {\rm i} \left( \Delta_c-U_0\sum_j\cos^2(k\bar x_j) \right) \bar a- \kappa \bar a + \eta \,, \label{eq:a eq}\\
 & \frac{\partial~}{\partial t} \bar x_j = \frac{\bar p_j}{m} \,, \label{eq:r eq} \\
 & \frac{\partial~}{\partial t} \bar p_j = - \partial_j V_{\rm ions} - U_0 \bar n \partial_j\cos^2(k\bar x_j) \,, \label{eq:p eq}
\end{align}
with $\bar n=|\bar a|^2$ and $\partial_j=\partial/\partial x_j$ is the gradient with respect to the spatial coordinate of the $j$-th particle (evaluated at the positions $\bar x_1,\ldots, \bar x_N$) . In deriving these equations, it is assumed that the field and the ions are in a separable state, which is valid in the mean-field regime. The stationary solution, obtained by setting $\partial_t\bar\zeta=0$, gives the ions' equilibrium positions in the mean-field limit. It also requires neglecting the fluctuations in the intracavity photon number, which is consistent when $\bar n>1$. 

We note that, after setting the time derivatives to 0 in eqs.~(\ref{eq:a eq}-\ref{eq:p eq}), and using the resulting relation $\bar a_{\rm eq}=\bar a_{\rm eq}(\bar x_1,\ldots,\bar x_N)$ in Eq. \eqref{eq:p eq}, the equilibrium conditions for the positions can be cast in the form
\begin{equation}
\label{equilibrium positions}
 \partial_j V_{\rm tot}\Bigl |_{\{\bar x_j\} }=0.
\end{equation}
Thus, in the mean-field limit the equilibrium positions of the ions minimize the effective total potential 
\begin{equation}
\label{V:tot}
V_{\rm tot}=V_{\rm ions}+V_{\rm eff}\,, 
\eeq
where the second term on the right-hand side is the mechanical potential given by \cite{Cormick:2012, Fischer}
\beq \label{eq:eff opt potential}
V_{\rm eff} = \frac{\hbar |\eta|^2}{\kappa} \arctan \left(-\frac{\Delta_{\rm eff}}{\kappa} \right)\,.
\eeq
In this expression, 
\beq
\Delta_{\rm eff} = \Delta_c-U_0\sum_j\cos^2(k\bar x_j)
\label{Delta_eff}
\eeq
is the effective detuning of the laser pump from the cavity field when the frequency shift of the cavity mode due to the atoms is taken into account.

For an ion chain in a harmonic potential, the interparticle distances are inhomogeneous and smaller at the center than at the edges of the chain. However, the central region is approximately homogeneous \cite{Morigi:2004}, and thus the distance $d_0$ at the chain center can be taken as the characteristic length scale of the ion chain. We assume that when the cavity is not pumped the interparticle distance $d_0$ is incommensurate with the cavity wavelength. In this case, in the presence of pumping and depending on whether the trap or the cavity forces dominate, the ion array can be found in one of two phases \cite{Garcia-Mata:2007, Benassi:2011, Fogarty:2015}. In the sliding phase, corresponding to low field intensities, the ions' equilibrium positions $\bar{x}_j$ are not correlated with the spatial shape of the cavity mode, so that the function $\cos^2(k\bar{x}_j)$ can take any value between 0 and 1, as in Fig.~\ref{schematic}~(a). In the pinned phase, reached at strong pumping, ions become confined within the wells of the optical lattice, as illustrated in Fig.~\ref{schematic}~(b) . The two phases are separated by a critical value of the potential $V_{\rm eff}$. This critical depth also separates two different dynamical behaviours, which we describe below. 

\subsection{Normal modes of crystal and field}\label{Sec:Opto}

We now consider the dynamics of the fluctuations of the ions and cavity variables about their corresponding mean-field values, assuming that the fluctuations are sufficiently small to truncate the expansion at second order. For this purpose we introduce the operators  $\delta x_j = x_j - \bar x_j$ and $\delta a = a-\bar a$ for the ions and the cavity field and solve the master equation:
\begin{equation}
\partial_t\varrho=\frac{1}{{\rm i}\hbar}[\delta H,\varrho]+{\mathcal L}\left[\delta a\right] \varrho\,,
\label{fluctuation evolution}
\end{equation}
where  ${\mathcal L}\left[\delta a\right]$ is quadratic in the cavity fluctuations and $\delta H$ is obtained expanding the Hamiltonian $H$ of \eq{eq:Htot} about the mean values of the cavity field and the ion positions. 
Explicitly we have: 
\be \delta H=\delta H_0 +\delta H_{\rm opto}\label{eq:deltaH},\ee
 with:
\begin{eqnarray}
&&\delta H_0=\sum_{j=1}\frac{p_j^2}{2m}-\hbar \Delta_{\rm eff}\delta a^\dagger \delta a +\frac{1}{2}\sum_{i,j}\frac{\partial^2V_{\rm ions}}{\partial x_i\partial x_j} \Bigl |_{\{\bar x_j\} } \delta x_i\delta x_j \nonumber\\
&& \quad\quad\quad -\hbar U_0 |\bar a|^2 k^2 \sum_j \cos(2k\bar x_j) \delta x_j^2 \,, \label{delta:0}\\
&&\delta H_{\rm opto}=-\hbar k U_0(\bar a^*\delta a+\bar a\delta a^\dagger)\sum_j \sin(2k\bar x_j)\delta x_j\label{delta:1}\,.
\end{eqnarray}

It is convenient to study the dynamics in terms of the normal modes of the ion chain, which diagonalize the Hamiltonian $\delta H_0$.  We label the normal modes by $\alpha=1,\ldots,N$, denote their frequency by $\omega_\alpha$ and define the associated phononic operators $b_{\alpha}$, which annihilate a quantum of vibration of energy $\hbar\omega_{\alpha}$. The modes are connected with the displacements $\delta x_j$ from the equilibrium positions $\bar x_j$ through the relation: 
\be
\delta x_j = \sum_{\alpha} \sigma_\alpha M_j^{\alpha}(b_\alpha+b_\alpha^\dagger)\,,
\label{normal mode coefs}
\eeq 
with $M_j^{\alpha}$ the orthonormal matrix diagonalizing the potential in Eq. \eqref{delta:0} and $\sigma_\alpha=\sqrt{\hbar/(2m\omega_\alpha)}$ the ground-state width of the oscillator corresponding to mode $\alpha$. In terms of normal-mode operators the Hamiltonian $\delta H$ of \eq{eq:deltaH} is given by \cite{Cormick:2012}:
\begin{eqnarray}
&&\delta H_0=\sum_{\alpha}\hbar \omega_{\alpha}b_{\alpha}^\dagger b_{\alpha}-\hbar \Delta_{\rm eff}\delta a^\dagger \delta a\,,
\label{H0}\\
&&\delta H_{\rm opto}=-\hbar\sum_{\alpha}(\chi_{\alpha}^*\delta a+\chi_{\alpha}\delta a^\dagger)(b_\alpha+b_\alpha^\dagger)\,.
\label{H:opto}
\end{eqnarray}
Here, we introduced the coupling coefficient $\chi_{\alpha}$ between phonon $\alpha$ and the cavity fluctuations, which reads:
\begin{equation}
\label{chi}
\chi_{\alpha}= \sqrt{\frac{\omega_R}{\omega_\alpha}}{\bar a}U_0\sum_{j}\sin(2k\bar x_j)M_j^{\alpha}\,.
\end{equation} 
In this expression, $\omega_R=\hbar k^2/(2m)$ is the recoil energy, scaling the mechanical effects of a cavity photon. In what follows we will consider regimes where $\omega_R\ll \omega_{\alpha}$, so that we can treat the optomechanical coupling between vibrations and cavity field fluctuations in perturbation theory. This regime corresponds to the so-called Lamb-Dicke regime, and is fulfilled when the spatial widths of the ions in the ground state are smaller than the light wavelength, namely, $k\sigma_\alpha=\sqrt{\omega_R/\omega_\alpha}\ll 1$.

\section{Cavity cooling of an ion chain}
\label{Sec:Cooling}

The Hamiltonian $\delta H_{\rm opto}$ couples the vibrational excitations of the chain with the cavity field fluctuations. The frequency of the external laser, $\omega_p$, can be tuned to drive the anti-Stokes resonance of a vibrational mode when the effective detuning of the pump photons from the cavity mode, as defined in \eq{Delta_eff}, obeys
\be\Delta_{\rm eff}\approx -\omega_{\alpha}\label{ResCond}.\ee 
The resulting processes can transfer an excitation from the phonon mode to the cavity mode as illustrated in Fig.~\ref{schematic}(e) and (f). The transfer can be irreversible if the photonic excitation is dissipated by cavity decay. Moreover, resolved-sideband excitation is reached when $\kappa<\omega_{\alpha}$. We note that the resonance condition \eqref{ResCond} is determined by a nonlinear equation, since $\Delta_{\rm eff}$ is a function of the ions' positions, which in turn also depend on the pump detuning with respect to the cavity, $\Delta_c$. 

In this section we discuss the prospects for using the scattering of cavity photons in order to simultaneously cool all axial vibrational modes to the ground state. We do this in a systematic way by starting in \seq{sec:analytics} with a simplified model that can be treated analytically and which allows one to identify the cooling characteristics in different parameter regimes, described in \seq{sec:CoolSliding} and \ref{parameters}. In \seq{Sec:CoolResults} we supplement this model with a thorough numerical study that demonstrates the possibility of simultaneous  cooling of all the axial modes to the ground-state using the cavity.

\subsection{Preliminary considerations}\label{sec:analytics}

We first analyze the cooling dynamics using a standard approach \cite{Stenholm:1986}, which is approximate but gives insight into the parameter regimes that are required in order to cool all the modes to the ground state. This treatment is valid for $\kappa \gg |\chi_{\alpha}|$, assuming that all vibrational modes are in the Lamb-Dicke regime and that photon fluctuations are close to their vacuum state at all times. In this regime, an excitation which is transferred from the vibrations to the cavity is then immediately damped by cavity losses. 

For convenience, we introduce the eigenstate $|n_\alpha\rangle$ with eigenvalue $n_{\alpha}$ of the number operator $b_\alpha^\dagger b_\alpha$ of mode $\alpha$. We further denote by $|0_c\rangle$ and $|1_c\rangle$  the zero- and one-photon states of the cavity mode fluctuations. The scattering processes which change the crystal motion are then either (i) cooling processes, characterized by the transition $|n_\alpha\rangle|0_c\rangle\to |n_\alpha-1\rangle|1_c\rangle\to |n_\alpha-1\rangle|0_c\rangle$, or (ii) heating processes, characterized by the transition $|n_\alpha\rangle|0_c\rangle\to |n_\alpha+1\rangle|1_c\rangle\to |n_\alpha+1\rangle|0_c\rangle$. In lowest order in the Lamb-Dicke regime each scattering process can change only the state of one mode at a time. The corresponding heating and cooling rates are obtained in second-order perturbation theory in the Lamb-Dicke parameter and are given by the expressions
\begin{equation}
\Gamma_{n\to n- 1}^\alpha=n_{\alpha}A_-^{\alpha}\,,
\end{equation}
and
\begin{equation}
\Gamma_{n\to n+ 1}^\alpha=(n_{\alpha}+1)A_+^{\alpha}\,,
\end{equation}
where
\begin{equation}
\label{heating-cooling rates}
A_\pm^{\alpha}=\frac{|\chi_{\alpha}|^2/\kappa}{1+(\Delta_{\rm eff}\mp\omega_{\alpha})^2/\kappa^2}\,.
\end{equation}

By means of these quantities one can write a set of rate equations for the dynamics of the excitations of mode $\alpha$. 
When $\chi_\alpha\neq 0$ and $A_+^{\alpha}<A_-^{\alpha}$, detailed balance applies and predicts a thermal stationary distribution with mean occupation number:
\begin{equation}
\label{n:steady}
n_{\alpha}^{\rm steady}=\frac{A_+^{\alpha}}{A_-^{\alpha}-A_+^{\alpha}}= \frac{(\Delta_{\rm eff}+\omega_{\alpha})^2+\kappa^2}{-4\omega_\alpha\Delta_{\rm eff}}\,.
\end{equation}
Within the regime of validity of this treatment, the cooling rate of mode $\alpha$ is given by $W_{\alpha}=A_-^\alpha-A_+^\alpha$. These considerations show that in principle the cavity can cool one or more modes of the chain to very low occupations $n_{\alpha}^{\rm steady}\ll 1$ when $\omega_{\alpha}\gg \kappa$ and $\Delta_{\rm eff}\approx-\omega_{\alpha}$, provided that $\chi_{\alpha}\neq 0$. The cooling of a mode is optimal when $n_{\alpha}^{\rm steady}$ of \eq{n:steady} is minimal, while at the same time the cooling rate as obtained from Eq.~\eqref{heating-cooling rates} is significantly faster than the competing thermalization processes due to the coupling to external environments.

\subsection{Cooling in the sliding phase}\label{sec:CoolSliding}

In the absence of the cavity field, the axial vibrational spectrum of the ion chain is composed of frequencies that vary between the lowest value $\omega_{\rm min}=\omega_t$, which corresponds to the trap frequency and is the oscillation frequency of the bulk, and the largest value $\omega_{\rm max}\sim\omega_0$, where
\beq
\label{omega:0}
\omega_0=\left(\frac{q^2}{4\pi\epsilon_0md_0^3}\right)^{1/2}\,,
\eeq
which is the frequency that scales the Coulomb interaction and gives the bandwidth of the axial mode spectrum. The parameter $d_0$ is the characteristic interparticle distance at the chain center in absence of the cavity mechanical forces, and is approximately given by \cite{Morigi:2004}
\beq
d_0\sim \left(\frac{q^2}{4\pi\epsilon_0 m\omega_t^2} \frac{3\log(N)}{N^2}\right)^{1/3}\,.
\eeq

For $N\gtrsim 10$, the inequality $\omega_0\gg \omega_t$ holds so that it is not possible to simultaneously fulfill the conditions $|\Delta_{\rm eff}+\omega_\alpha|\ll \kappa$ and $\omega_{\alpha}\gg \kappa$ for all modes. This situation is maintained for non-zero, but small, intensities of the laser pumping the cavity, such that the ions are found in the sliding phase. As an example, in Fig.~\ref{schematic}(c) we show the output spectrum of the normal modes, which is defined in Eq.~\eqref{Eq:Snu} of the Appendix, for low pump intensity. In this case one can resonantly drive the anti-Stokes sideband of one, or at most a few, modes at a time (see Fig.~\ref{schematic}(e)). However, other modes can be simultaneously heated up in the process. In addition, some of the coupling strengths $\chi_{\alpha}$, and thus the corresponding cooling rates, typically become very small. This is easily seen when considering sideband cooling of the bulk (center-of-mass) mode. Then, $M_j^\alpha=1/\sqrt{N}$ and the cooling rate at resonance is 
\begin{equation}
W_{\rm bulk}\simeq \frac{\omega_R}{\omega_t}\frac{U_0^2|\bar a|^2}{\kappa}\frac{\left[\sum_j\sin(2k\bar x_j)\right]^2}{N}\,.
\end{equation}
When the cavity wavelength $\lambda$ is incommensurate with $d_0$ and for a large number of ions, one finds $[\sum_j\sin(2k\bar x_j)]/N\to 0$, such that the cooling rate tends to vanish with increasing numbers of ions.

\subsection{Parameter regime for ground-state cooling using frustration}\label{parameters}

The scenario described above can change quite dramatically if we consider the regime where the cavity mechanical potential appreciably modifies the crystalline ground-state structure. When the pump exceeds the critical value of the sliding-pinned transition, the ions of the array get trapped in the wells of the optical potential, approaching the minima of this optical lattice as the pump intensity is increased. This leads to a bunching of the vibrational mode frequencies about the oscillation frequency determined by the cavity optical lattice, as is illustrated in the spectrum shown in Fig.~\ref{schematic}~(d).

\begin{figure}[h!]
\includegraphics[width=0.99\columnwidth]{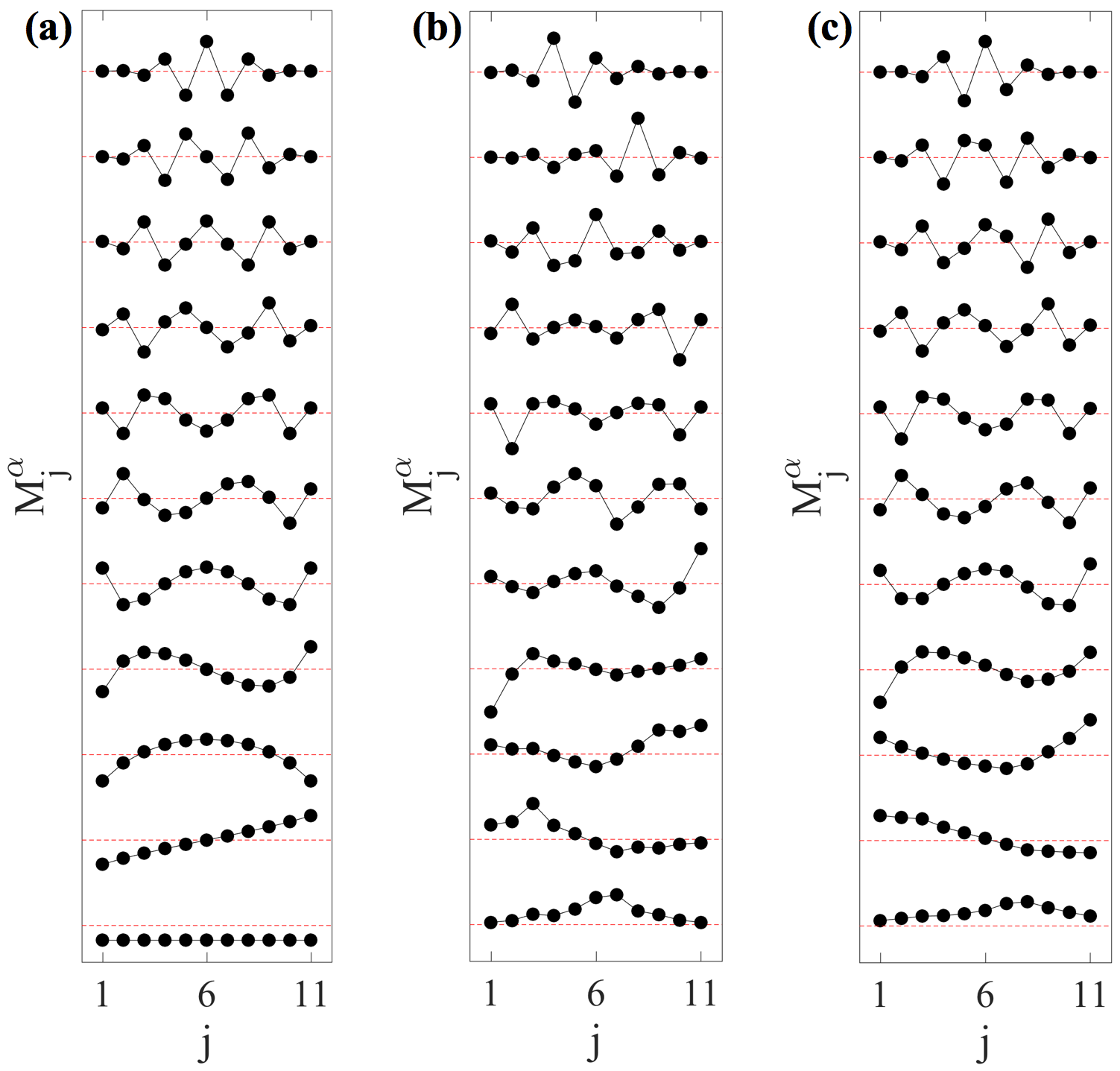}
\caption{Axial displacements of the ions for each normal mode~$\alpha$, for parameters such that the chain is (a) in the sliding phase, (b) in the pinned phase, close to the sliding-to-pinned transition and (c) deep in the pinned phase (modes are ordered from the lowest frequency at the bottom to the highest at the top). The displacements are given by the values of the matrix $M_j^\alpha$ in Eq. (\ref{normal mode coefs}) as a function of the ion $j$ for 11 ions, where $j$ labels the ions from left to right. The red horizontal lines indicate the value of zero axial displacement (with respect to the equilibrium position of each ion), about which the corresponding value of $M_j^\alpha$ is reported. Subplots (a) and (c) have been calculated for the same parameters of the cavity output spectra in Fig. \ref{schematic}(c) and (d), with $\eta=1.5\kappa$ and $\eta=300\kappa$ respectively, while subplot (b) is with $\eta=55\kappa$. The presence of the kink in panels (b)-(c) significanlty modifies the normal modes.\label{Fig:M} 
}
\end{figure}

This leads to two relevant features for cooling: (i) the bandwidth $\delta\omega$ of the vibrational spectrum becomes narrower and the central frequency $\bar \omega$ higher, so that there exists a value of the pump strength $\eta$ above which $\delta\omega\ll\kappa$ and $\kappa\ll\bar\omega$ at the same time, which makes it possible to address the anti-Stokes sideband of all modes simultaneously (see Fig.~\ref{schematic}~(f)).  (ii)  Due to the inherent frustration in this system, even for a large depth of the optical potential the equilibrium configuration contains some ions positioned away from the minima, which significantly modifies the spatial structure of the vibrational modes as illustrated in Fig.~\ref{Fig:M}. These ions are defects in the structure of the chain and are known as kinks. The existence of kinks in this system is an essential factor to achieve cavity cooling, as the kinks strongly couple to the cavity field, providing simultaneous cooling of all modes. The use of this coupling to achieve ground-state cooling of the whole chain will be analyzed in detail in the following subsection.

To give an estimate of the relevant parameters, we note that the regime of interest requires that the optical potential is deep enough to pin most ions near the lattice minima. Therefore minimization of the total effective potential corresponds approximately to a minimization of the optical potential, overcoming the increase in Coulomb energy. Since the Coulomb potential is proportional to $1/d_0$, the required depth of the optical potential will decrease as interparticle distances are increased. To estimate the required optical depth we consider an ion that must be shifted from an initial position at a maximum of the optical potential to a point near a minimum, which means a displacement of approximately $\lambda/4$. To simplify the calculation, we consider only the Coulomb potential associated with the nearest neighbours, whose positions are assumed to be fixed. For $^{174}$Yb$^+$ ions in a Paul trap overlapped with a cavity of wavelength $\lambda\simeq369$~nm, linewidth $\kappa=2\pi \times 0.2$~MHz, and $U_0 = 0.5 \kappa$, we choose $d_0\simeq 6.8\mu$m. Then, the change in Coulomb energy associated with a $\lambda/4$ displacement is comparable with the change in optical potential if $U_0\langle n\rangle\simeq2\pi\times 1$~GHz. This in turn corresponds to about $\langle n\rangle\simeq10^4 $ photons and hence $\eta\simeq 100\kappa$ (the actual value will depend also on the detuning $\Delta_c$). This estimate is consistent with the assumptions at the basis of the optomechanical Hamiltonian provided that the atom-field detuning is sufficiently large, so we take $\Delta_d=2\pi \times 12$~GHz to be within the regime of validity of our approximations. We further note that we choose $\Delta_d>0$, and thus $U_0>0$; therefore, the minima of the optical potential are at the nodes of the field. Our examples will focus on the case of $N=11$ ions, with an axial trap frequency $\omega_t\simeq 2\pi\times 100$kHz.


\subsection{Cooling rates and mean vibrational number at steady state}\label{Sec:CoolResults}

\begin{figure}[h!]
 \includegraphics[width=0.9\columnwidth]{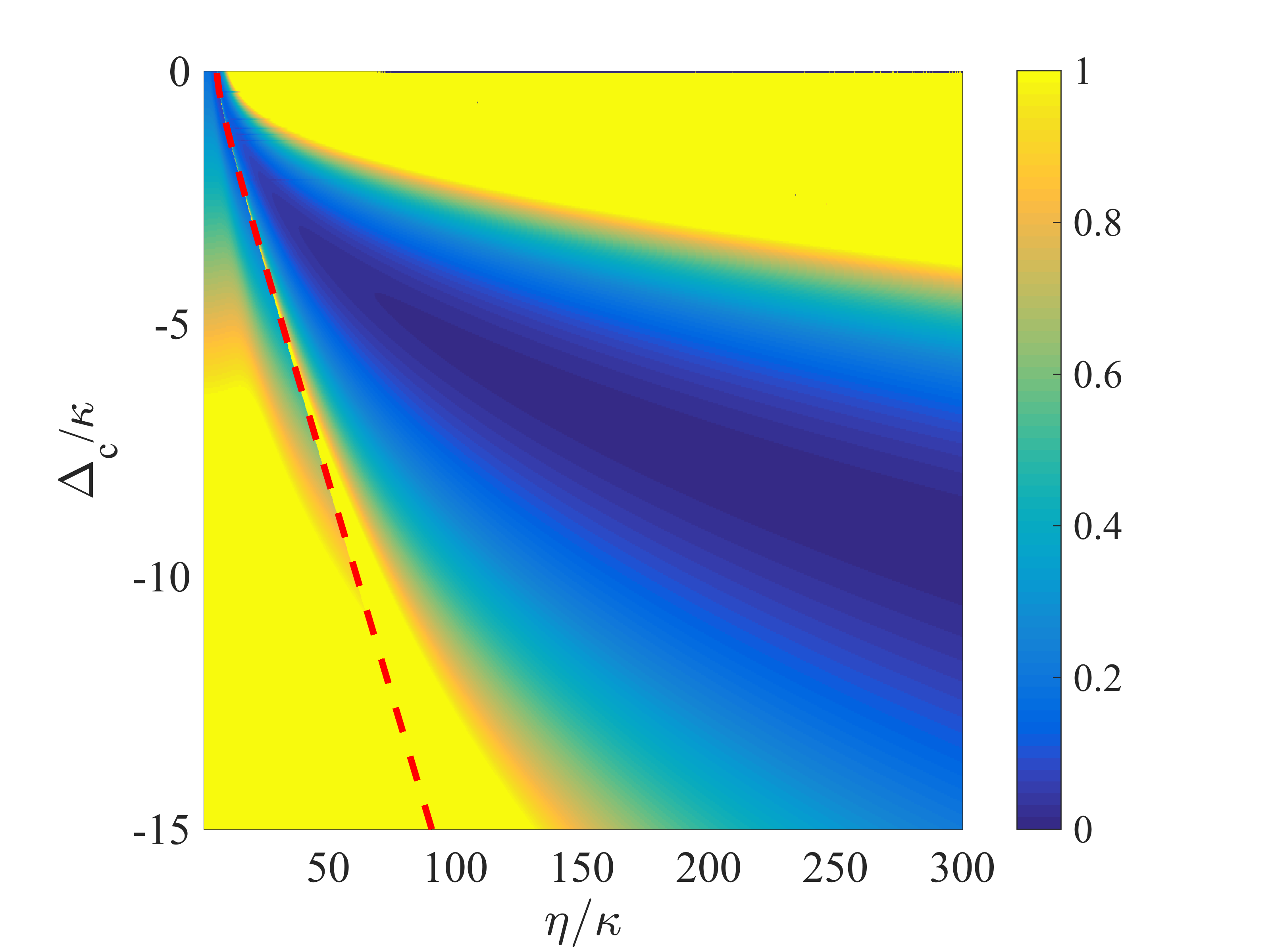}
\caption{Mean asymptotic excitation number $\bar n_\alpha$, averaged over all modes with non-zero coupling to the cavity fluctuations, for $N=11$ ions (see text for the relevant parameters). The dashed line indicates the sliding-pinned transition, to the left of which the symmetric modes of the chain are decoupled from the cavity. Phonon numbers above 1 are indicated by the brightest color in the figure.\label{mean nbar} 
}
\end{figure}

To study the effectiveness of cavity cooling on the ion chain we numerically examine the mean excitation number ${n}^{\rm steady}_\alpha$ of each vibrational mode that is coupled to the cavity. The stationary mode occupation is evaluated by calculating the expectation value of the number operator $b_{\alpha}^\dagger b_{\alpha}$ over the stationary density matrix $\varrho^{\rm steady}$ as given in \eq{Eqnsteady} of the Appendix. 
Fig.~\ref{mean nbar} shows $ n_{\alpha}^{\rm steady}$ averaged over all modes that have non-zero coupling with the cavity fluctuations, for varying $\Delta_c$ and $\eta$. The dashed line indicates the sliding-pinned transition. Due to the symmetric arrangement considered, in the sliding phase half of the modes are not coupled with cavity fluctuations and have therefore no well-defined asymptotic state \cite{Fogarty:2015}; these modes are not included in the region of the plot to the left of the dashed line. After the transition, in the pinned phase, the symmetry breaking couples all of the vibrational modes with the photonic fluctuations. We note that the symmetry in the sliding phase can be easily lifted by shifting the cavity with respect to the trap and is not an essential feature of our model. 

It is clear from Fig.~\ref{mean nbar} that, within the pinned phase, the parameter region where low excitation numbers are reached corresponds to a lower value of $\Delta_c$ as $\eta$ is increased. This is because, as formula (\ref{n:steady}) shows, the coldest asymptotic states are found when the cavity field is close to resonance with the vibrational modes, and the mode frequencies increase with increasing $\eta$. However, in the region where $\eta$ is very large the coupling strengths $|\chi_{\alpha}|$ between the vibrational modes and the cavity fluctuations, given in Eq. \eqref{chi}, tend to become very small as the ions become localized closer to the minima of the optical lattice, which is detrimental for the cooling rates, estimated in Eq. \eqref{heating-cooling rates}. Choosing a cavity detuning of $\Delta_c= -8.5\kappa$ the coupling strengths $|\chi_{\alpha}|$ are plotted as a function of $\eta$ in Fig.~\ref{mode_spec}~(a). The disparities in the coupling strengths for different modes result from the dependence of $\chi_{\alpha}$ on both the phase of the ions' equilibrium positions within the cavity mode, and each mode's spatial profile, shown in Fig.~\ref{Fig:M} for a few values of pump strength $\eta$. 

\begin{figure}[h!]%
\includegraphics[width=\columnwidth]{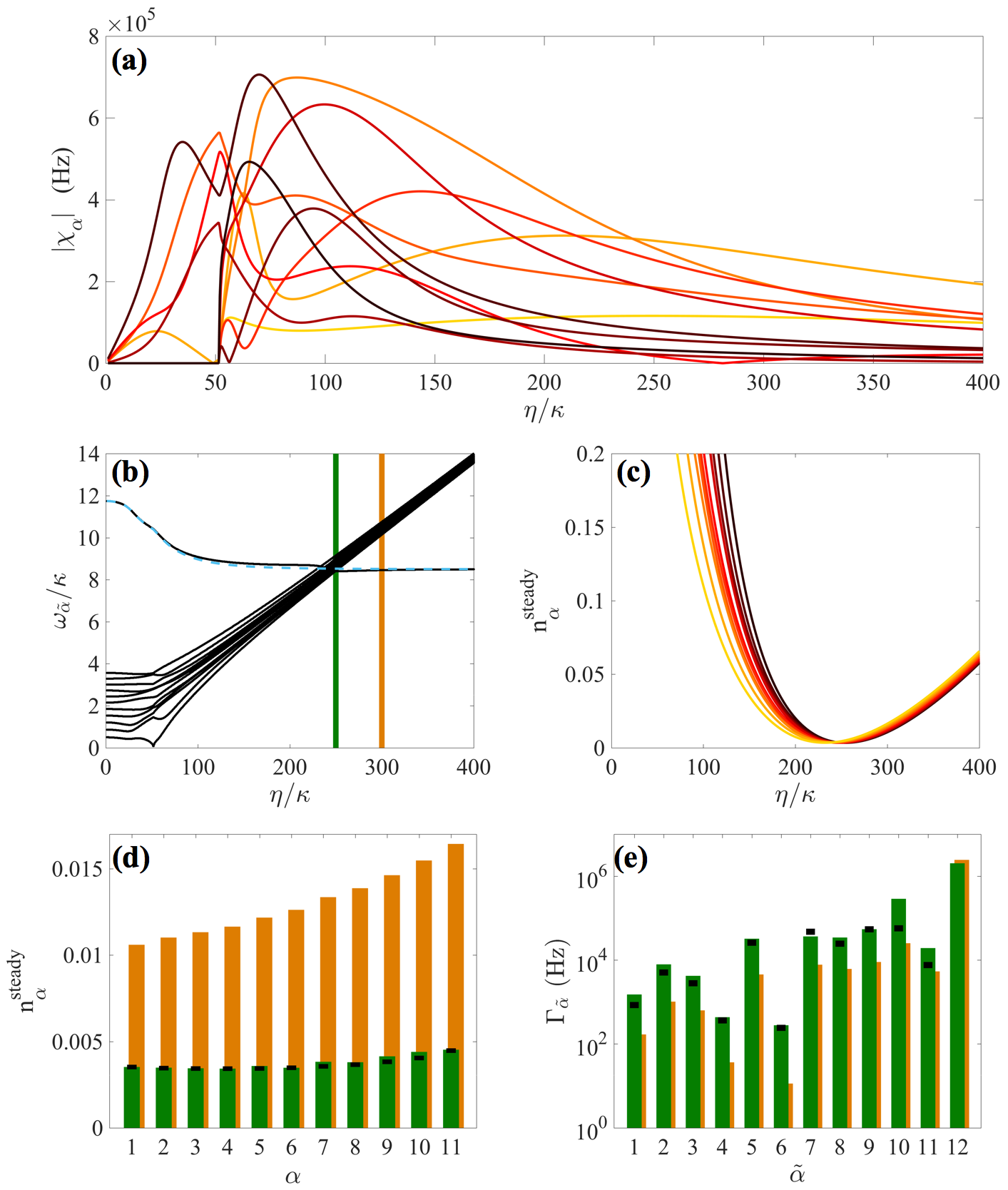}
\caption{Analysis of cavity cooling for $\Delta_c=-8.5\kappa$ (the remaining parameters are given in the text). (a) Coupling coefficients $|\chi_{\alpha}|$ between each phonon mode $\alpha$ and cavity fluctuations as a function of $\eta$, ordered from lowest frequency mode (black line) to highest frequency mode (yellow). (b) Frequencies of the generalized phonon-photon modes as a function of the pump strength. For this detuning there is a resonance between the photon fluctuation frequency $-\Delta_{\rm eff}$ (blue dashed line) and the vibrational modes at $\eta=250\kappa$ (green line); we investigate the cooling at this point and at $\eta=300\kappa$ (orange line). (c) Mean steady-state occupation of each vibrational chain mode as a function of $\eta$ and (d) at $\eta=250\kappa$ (green bars) with black lines representing the analytic results from Eq.~\eqref{n:steady}. The orange bars are at a $20\%$ fluctuation of the pump strength to $\eta=300\kappa$. (e) Cooling rates $\Gamma_{\tilde{\alpha}}$  (in log scale) for each of the generalized modes, for the same values of $\eta$ as in (d). For these parameters, the last eigenmode has a dominantly photonic character, while each of the first 11 generalized modes is composed mainly by a particular vibrational mode that is mixed with cavity fluctuations. The black lines indicate the analytic estimates of the cooling rates from Eq. \eqref{heating-cooling rates} for the corresponding vibrational modes. Panels (d) and (e) demonstrate that the chain can be robustly cooled close to its ground state for all modes simultaneously.
}
   \label{mode_spec}
\end{figure}

To study in detail the phonon-photon coupling and cooling rates, the equations of motion for the coupled cavity and chain fluctuations are rewritten in a compact form of a linear system (see \eqref{eq:M} and \cite{Cormick:2013}) starting from Eq.~\eqref{fluctuation evolution} and 
Eqs.~\eqref{H0}-\eqref{H:opto}. The diagonalization of this system leads to generalized eigenmodes with corresponding eigenvalues $\lambda_{\tilde{\alpha}}$, where we use $\tilde \alpha$ to stress that we refer to a mode mixing phononic and photonic character. The cooling rates are obtained from the real part of these eigenvalues, 
\be
\label{cool:rate}
\Gamma_{\tilde \alpha} = -2 {\rm Re} (\lambda_{\tilde \alpha} )\,.
\eeq
The resulting rates $\Gamma_{\tilde \alpha}$ sum up to $2\kappa$ and include the damping of the cavity fluctuations. We analyze the regime in which the vibrational modes are cooled and so we restrict to parameter values for which all $\Gamma_{\tilde \alpha}$ are non-negative, which requires $\Delta_{\rm eff} < 0$ \cite{Cormick:2013}. Naturally, the eigenfrequencies of the generalized modes are given by the imaginary part of the eigenvalues
\be
\label{eigen:freq}
\omega_{\tilde \alpha} = |{\rm Im} (\lambda_{\tilde \alpha} )|\,.
\eeq

Fig. \ref{mode_spec}~(b) shows the frequencies of the system's generalized eigenmodes, Eq.~(\ref{eigen:freq}), as a function of the pump strength $\eta$, for the same detuning as in Fig.~\ref{mode_spec}~(a). For $\eta\simeq 0$, the eigenmodes are approximately the same as the modes of the decoupled fluctuations,  namely, vibrational normal modes of the ion chain, and a cavity fluctuation mode. In this limit, photon fluctuations have an associated frequency $\Delta_{\rm eff}\simeq\Delta_c-U_0N/2$, because ions are distributed over all values of the mode spatial function $\cos(kx)$. At $\eta \simeq 50 \kappa$ we identify the sliding-to-pinned transition point, for which the lowest mode frequency vanishes. The regime close to the transition point will be discussed in Sec.~\ref{Sec:CoolKink}. To the left of the transition the chain is in the sliding phase, while as $\eta$ is increased, most ions become trapped closer to the nodes of the optical potential. The normal modes then become increasingly bunched about the frequency value of the lowest lattice band, the bandwidth being determined by the kink density. For our choice of parameters, the resonance between the motional modes and the cavity fluctuations occurs at $\eta\simeq 250\kappa$. For larger $\eta$, the normal modes of the chain increase their frequency moving away from the resonance with the cavity field fluctuations, whose frequency tends towards the value $\Delta_{\rm eff}\simeq \Delta_c$.

Fig. \ref{mode_spec}~(c) displays the stationary mean occupation of each of the ions' vibrational modes as a function of $\eta$ (an explanation of the calculation is given in the Appendix). For small $\eta$ the individual occupations possess a broad spread in values due to the large bandwidth of the generalized modes in the sliding phase. As $\eta$ is increased these occupations begin to bunch and reach a minimum at the point of the resonance where all anti-Stokes transitions are resonantly driven. Panel (d) shows the stationary mean occupation of each vibrational mode at resonance (green bars), whereby all the modes are close to the ground state. These results are in very good agreement with our approximate expression Eq.~\eqref{n:steady}, represented as the black lines. The orange bars correspond to the variations due to a fluctuation of the pump frequency by 20\% off the optimal value. This shows that even when the vibrational modes moves out of resonance with the cavity field the effect of the optomechanical cooling is still strong. 

The cooling rates of the generalized modes $\Gamma_{\tilde{\alpha}}$, calculated from Eq. \eqref{cool:rate}, are plotted in Fig.~\ref{mode_spec}~(e) for $\eta=250\kappa$ and $\eta=300\kappa$. The modes $\tilde{\alpha}=\{1,\dots,11\}$ correspond mostly to motional modes of the chain, although there is some mixing with the cavity fluctuations. The rate for $\tilde{\alpha}=12$ is approximately equal to the damping rate of the cavity field. The black lines identify the corresponding analytic cooling rates for the vibrational modes as estimated in Section~\ref{sec:analytics}. For these parameters, the ion chain can be cooled to the ground state on tens of milliseconds time-scales for all the modes.

\begin{figure}[h!]%
  \includegraphics[width=0.99\columnwidth]{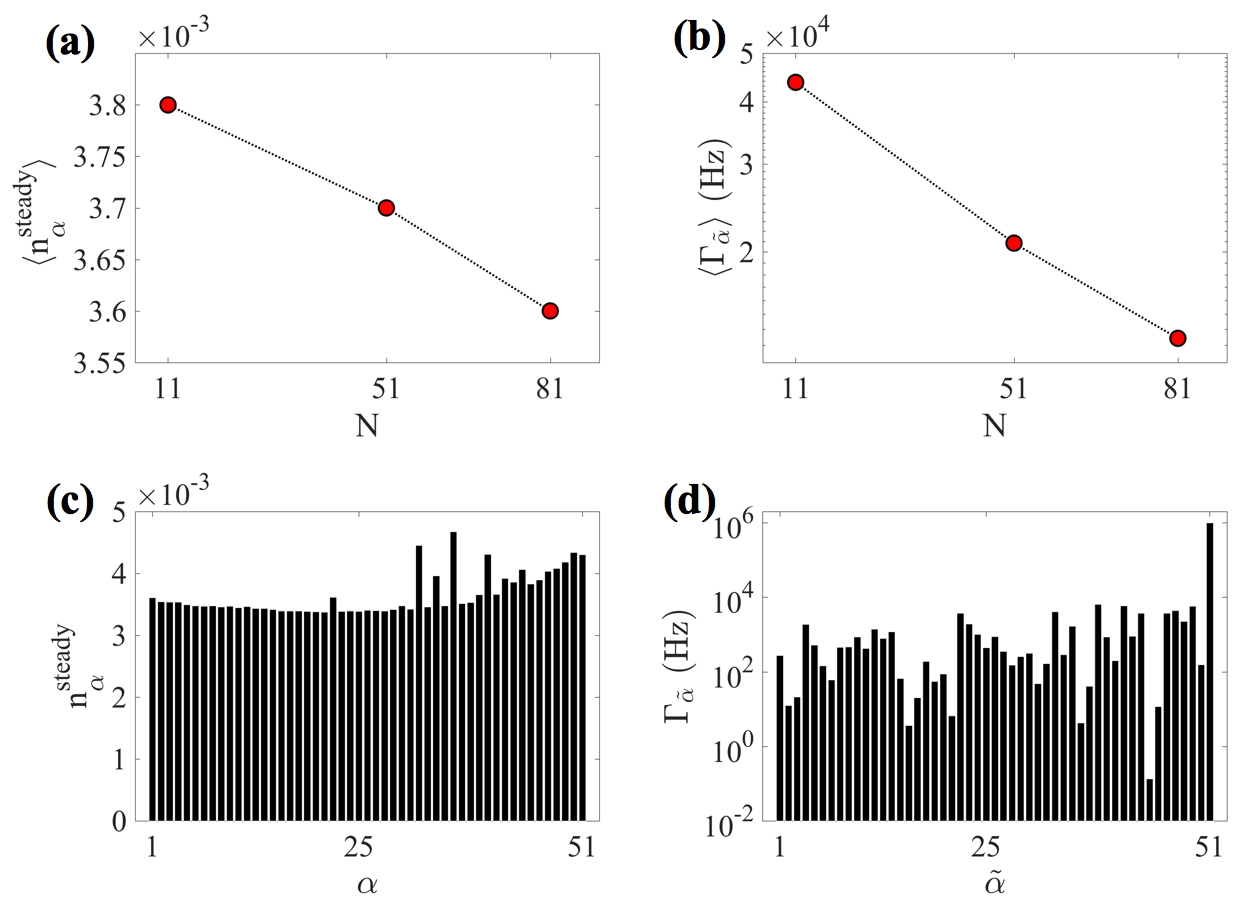}%
\caption{(a) Mean steady-state occupation averaged over all vibrational modes as a function of the chain size, for $N=11, 51, 81$. For each $N$ the distance between the ions at the chain center and the cavity potential depth are kept constant. (b) Average cooling rates (in Hz) of the generalized modes as a function of the chain size (in logarithmic scale for the vertical axis). (c) Steady state occupation of the vibrational modes for $N=51$, and (d) corresponding cooling rates of the 51 eigenmodes which are dominantly vibrational (in logarithmic scale). We note that for studying the scaling behavior we fix $\eta$ and $\kappa$ so that the results are not optimized for the larger system size (see the text for parameters and discussion).
}
\label{nsteadyN}
\end{figure}

To investigate cooling of larger chains, in Fig.~\ref{nsteadyN}(a) the average of the ion mode occupation numbers are shown as a function of chain size $N$ at $\eta=250\kappa$ and $\Delta_c=-8.5\kappa$. In Fig.~\ref{nsteadyN}(b) the mean cooling rates of the generalized modes are plotted, with the $\tilde{\alpha}=N+1$ mode omitted as it predominantly describes the damping of the cavity fluctuations. We find a slight decreasing trend in the mean occupation accompanied by a stronger dependence of the average cooling rate on the number of ions (we note that $\kappa$ is kept fixed while the number of modes that are to be cooled increases $\propto N$). The reason why the mean cooling rate does not decrease $\propto N^{-1}$ is the scaling of the trap frequency $\omega_N \propto \sqrt{\log N}/N$. Thus, the ion separation at the center of the trap is kept fixed, which leads to a density of kinks along the chain that stays approximately constant as $N$ is increased \cite{Fogarty:2015}. To understand why we still see a decrease in the cooling rate, we examine in detail the mean occupation numbers and the cooling rates for each mode of the $N=51$ chain in panels (c-d). In general the cooling rates for the majority of the modes are high, resulting in cooling times of tens of milliseconds, however a small number of weakly coupled modes have lower cooling rates. As a result of the deformation of the individual modes by the cavity potential (see Fig.~\ref{Fig:M}) a mode can become almost decoupled from the cavity ($\chi_{\alpha}\approx0$) for a specific value of the cavity depth. As the number of modes of the chain increases with $N$, the occurrence of a decoupled mode becomes more probable, whereby optimization of the cavity depth must be carried out to avoid such cases. However, in Fig.~\ref{nsteadyN} we take a fixed cavity depth which is therefore not optimized for the larger system sizes shown. 

\section{Selective manipulation of a kink}\label{Sec:CoolKink}

While deep in the pinned phase the normal mode spectrum becomes bunched, close to the sliding-pinned transition the lowest frequency mode spectrally separates from the rest of the vibrational frequencies, as visible in Fig. \ref{mode_spec}~(b). Exactly at the transition point, the frequency of the lowest motional mode vanishes \cite{Braun}. About this transition, if the laser pump is properly tuned, the kink mode can be individually cooled to the ground state by the cavity fluctuations. Moreover, the light emitted at the cavity output provides a natural way to perform spectroscopy of the kink. 

\begin{figure}[ht!]%
 \includegraphics[width=0.99\columnwidth]{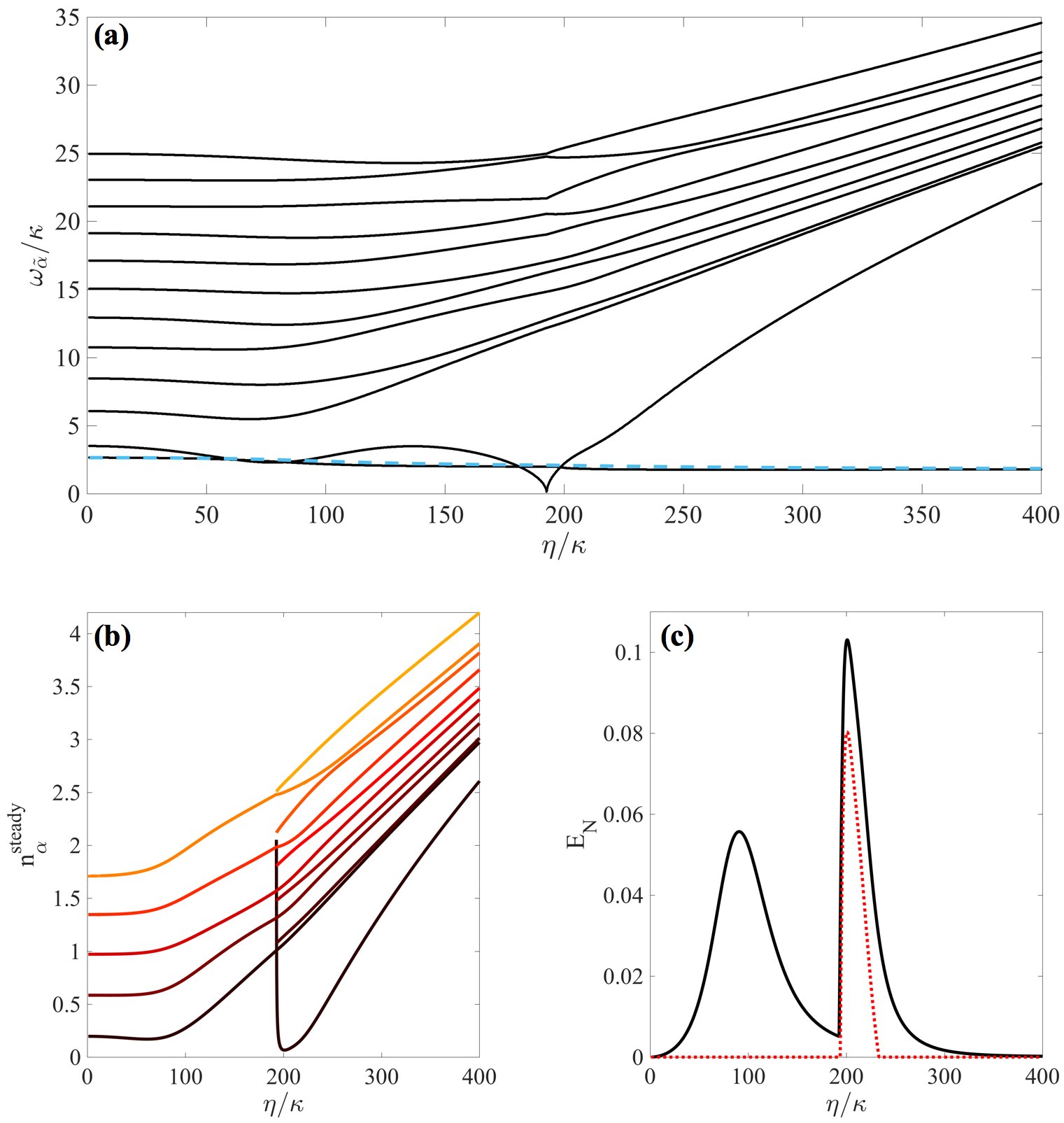}%
\caption{(a) Generalized eigenfrequencies $\omega_{\tilde{\alpha}}$ in units of $\kappa$, when fixing $\Delta_c=-1.8\kappa$ so that the cavity is close to resonance with the kink mode at $\eta\simeq 200\kappa$. The blue dashed line is the effective cavity detuning $-\Delta_{\rm eff}/\kappa$. (b) Asymptotic occupation numbers of the vibrational modes; only modes with nonzero coupling to the cavity are shown. The kink mode is shown in black. (c) Entanglement between the cavity fluctuations and all the modes (black solid line) and between the cavity and the kink mode only (red dashed line), quantified by the logarithmic negativity. Parameters are given in the text.
}
\label{kink_modes}
\end{figure}

The physical nature of the kink mode is best understood when the center of the ion trap coincides with a maximum of the optical potential, as is the case for our choice of the total Hamiltonian in \eq{eq:Htot}. In this case, in the sliding phase and for an odd number of ions, the central ion of the chain is located at a maximum of the cavity potential. As the pump strength is increased, this ion remains in the same place while the others slowly tend to occupy positions closer to minima of the optical potential. Eventually, a critical point is reached such that the ground state is no longer symmetric, because the central ion is shifted away from the center. After this point the chain ground state corresponds to the pinned phase \cite{Klafter:1990}. The minimization of the total potential then comes at the expense of an increase of the Coulomb repulsion, and the chain exhibits at its center a structural defect, or kink, since the interparticle distances become irregular. 

In this situation, the lowest vibrational mode, which we call the kink mode, is localized at the center, as can be seen by inspecting the spatial shapes of the motional modes in Fig.~\ref{Fig:M} (b). Close to the transition point the frequency gap between the kink mode and the next lowest frequency excitation of the spectrum is approximately given by the local frequency of the optical lattice wells. In the vicinity of the transition, this frequency is of the order of $\omega_0$, Eq. \eqref{omega:0}. Therefore, the kink mode can be better spectrally resolved the larger $\omega_0$ is, i.e.~the smaller is the distance between the ions at the chain center. 

With this property in mind, in the following analysis we take the same cavity and ions parameters as in Sec.~\ref{Sec:Cooling}, but choose a tighter trap frequency of $\omega_t=2\pi\times 700$~kHz resulting in an interparticle distance at the trap center of approximately 1.9~$\mu$m in absence of the optical potential. Selective coupling with the kink mode can be achieved by setting the pump detuning to be $\Delta_c=-1.8\kappa$. This optimal value of the detuning is found by considering the nonlinear shift of the cavity frequency, Eq. \eqref{Delta_eff}, and the fact that the cavity mode strongly couples with the kink mode. For these parameters, the sliding-pinned transition takes place at $\eta\simeq190\kappa$ and the detuning is such that the cavity mode becomes resonant with the kink mode just after the transition, at $\eta\simeq200\kappa$. This resonance is visualized as the crossing of the two lowest modes in Fig. \ref{kink_modes}~(a), which shows the spectrum of the generalized photon-phonon modes as a function of the pump strength (where the blue dashed line is $-\Delta_{\rm eff}/\kappa$).  The resonant coupling ensures that the kink mode is cooled close to its zero-point motion, while the other modes are off resonance and have higher asymptotic occupation numbers, as plotted in Fig.~\ref{kink_modes}~(b). 

Another result of this coupling is the creation of entanglement between the kink and the cavity fluctuations. In Fig.~\ref{kink_modes}~(c) we plot the entanglement between the cavity and the whole collection of vibrational modes, and also the entanglement between cavity and kink mode, in terms of the logarithmic negativity \cite{logneg} (more details are given in the Appendix). In the sliding phase, in which there is no kink mode, the cavity mode is correlated with half of the motional modes, while the rest are decoupled because of the symmetry. In particular, the level crossings seen in Fig. \ref{kink_modes}~(a) for $\eta$ between 50 and 100 do not lead to entanglement between the cavity fluctuations and the lowest mode since in the sliding phase this mode is decoupled from the cavity fluctuations. The local maximum of the logarithmic negativity found at $\eta\simeq 100\kappa$ is due to strong coupling between the second-lowest phonon mode and the cavity field in this region. Within the pinned phase, at the point of the resonance between the kink mode and the cavity fluctuations a spike in the logarithmic negativity is observed. The plot clearly indicates that the majority of the photon-phonon correlations in this regime is associated with the kink and not with the rest of the vibrational modes.

\begin{figure}[h!]%
 \includegraphics[width=0.99\columnwidth]{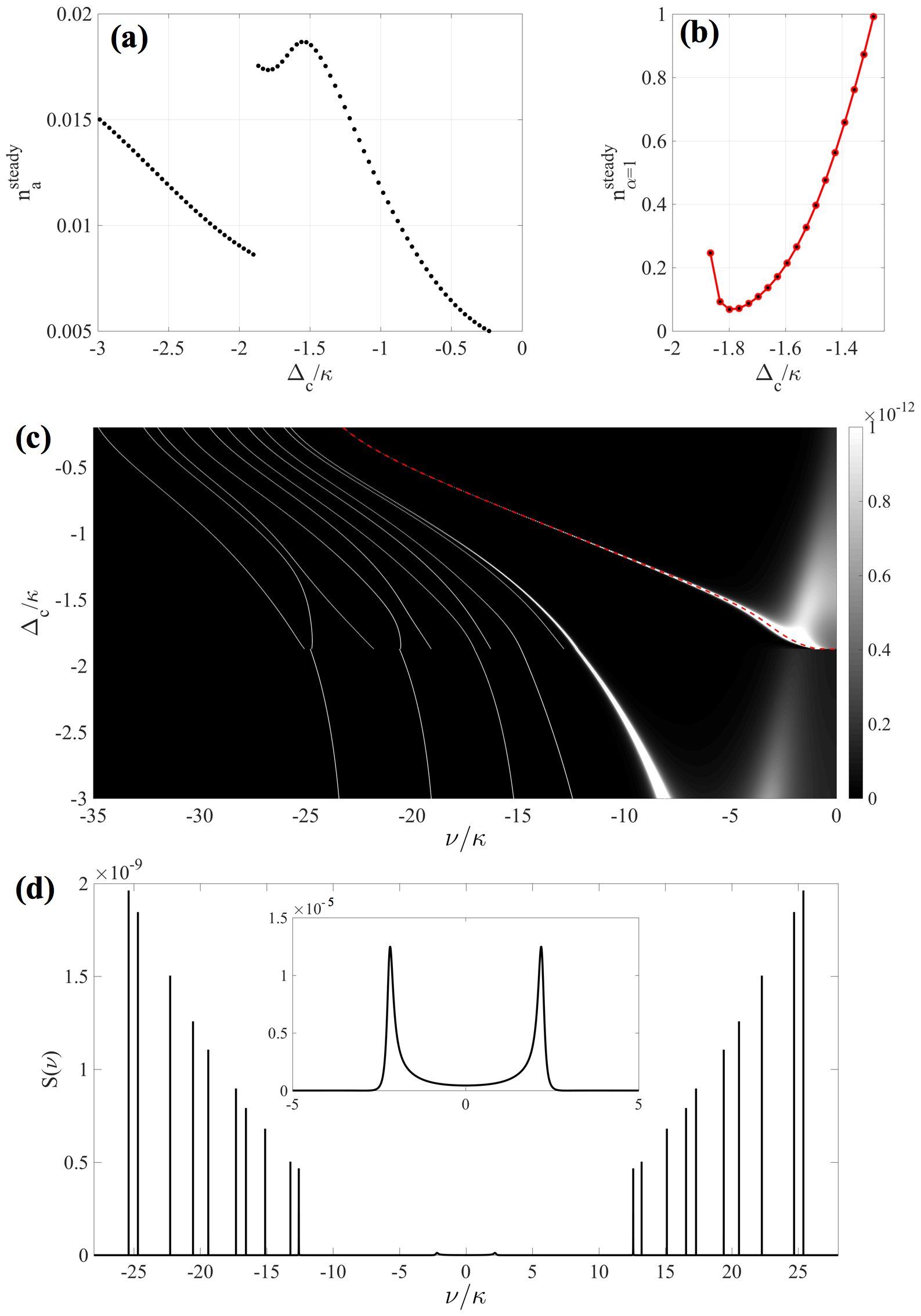}%
\caption{(a) Mean value of number of photon fluctuations as a function of the cavity detuning and (b) the mean occupation of the kink mode, which is minimum at resonance; the line is interrupted in the range in which the kink mode is not coupled to the cavity. (c) Spectrum at the cavity output (in units of $1/$Hz) as a function of the cavity detuning and the photon frequency shift $\nu$ (in units of $\kappa$). The peak due to the kink mode vanishes near $\nu=0$ at $\Delta_c\approx-1.9\kappa$ and is highlighted by the red dashed line. There is a cutoff in the color scale at $10^{-12}$Hz$^{-1}$ to enhance the spectrum of the kink mode. (d) Output spectrum at $\Delta_c=-1.8\kappa$, when the kink is resonantly coupled to the cavity mode; the inset shows a magnification of the kink mode which is spectrally separated by a gap.
}
\label{kink_spc}
\end{figure}

We now investigate the ability to spectrally resolve this kink mode. In this case it is convenient to hold the pump strength constant at $\eta=200\kappa$ and vary the frequency of the pump laser $\omega_p$ around the resonant value of $\Delta_c\approx-1.9\kappa$. The resonance point, and thus the location of ground-state cooling of the kink, can be extracted from the mean photon number fluctuations, shown in Fig.~\ref{kink_spc}~(a) (details about the calculation are given in the Appendix). At the point at which the kink mode is cooled to its minimum value (see panel (b)) there is a local minimum of the intensity. This is caused by a simultaneous broadening and diminishment of the kink spectral peak when it is cooled to its ground state. Shifting the detuning to larger negative values decouples the kink mode from the cavity mode, resulting in a discontinuous drop in the intensity at the sliding-pinned transition. This decoupling can also be identified as a disappearance of the kink peak at $\Delta_c\approx-1.9\kappa$ in Fig.~\ref{kink_spc}~(c), which shows the spectrum of the cavity fluctuations as a function of the photon frequency shift $\nu$ and the cavity detuning $\Delta_c$. The kink mode is highlighted by the red dashed line, and it can be seen to be significantly broadened about the point of resonance and ground-state cooling. The $\tilde{\alpha}=2$ mode has also a strong mixing with the cavity field and experiences spectral broadening, whereas higher energy modes are far off resonance with the cavity field and exhibit narrow linewidths. In Fig.~\ref{kink_spc}~(d) the full output spectrum is plotted for $\Delta_c=-1.8\kappa$, where at the point of the resonance the kink mode is well separated from the bulk modes and can be spectrally resolved (see inset).

\section{Conclusions}\label{Sec:TheEnd}

Optomechanical coupling to a high-finesse resonator dynamically modifies the vibrational spectrum of an ion chain and can induce scattering processes with a high cross-section. As a result of bringing all the modes simultaneously to resonance with the cavity mode, the entire chain can be cooled to the zero-point motion. We have calculated the final mode occupations and cooling rates for state-of-the-art setups \cite{Cetina:2013,Vuletic:2015}, and shown that mean phonon occupations on the order of $10^{-2}$ can be achieved in tens of milliseconds for tens of ions. 

The key aspect that permits significant cooling rates of all bunched modes simultaneously is keeping a fixed density of kinks in the chain. The kinks prevent the particles from organizing in a regular lattice that is commensurate with the cavity wavelength, a structure that would have a low scattering rate of photons as the particles occupy the extrema of the cavity mode profile.

In long chains, as the chain size is increased it may still happen that some modes become decoupled from the cavity so that the cooling time of these modes increases. This problem could be at least partially solved by embedding ions into the chain which are impurity defects and that selectively couple with the optical field \cite{Fogarty:2013,Lin:2013,Schmiedt:2015}. Moreover mass defect ions interact with the kinks and lead to even richer possibilities of stabilizing kinks and changing the normal modes \cite{Braun,solitons13,solitons13b}, which are intriguing phenomena on their own. From a practical point of view, the fastest implementation of a cooling protocol leading to a final vibrational ground state could involve a sweep in the parameter space given by $\Delta_c$ and $\eta$. This would allow one to start with higher cooling rates and adjust the parameters toward regions with lower excitation numbers, a useful strategy in resolved sideband cooling \cite{Eschner:2003}.

The ideas that we discuss in this work could be extended to planar or three-dimensional crystals, and furthermore serve as inspiration for developing laser cooling schemes of solids \cite{Dunlop,Sheik}, where the coupling to the resonator is being considered in order to increase the cooling rates \cite{Sheik,Solid:Cavity}. Here, one could make instrumental use of the mode shaping induced by static friction in order to modify the mode spectrum and thus cool a larger band of modes. 

Our study, in addition, shows how optomechanical coupling to a cavity mode can enable the manipulation of topological defects (known as solitons, kinks or dislocations) in a nonlinear system of oscillators in the quantum regime \cite{solitons08,solitons10,landa}. The coupling of the particles to the cavity allows one to achieve nonclassical states of quasi-particles -- localized modes of the stationary state of a large and noisy system. This represents a novel instance of quantum reservoir engineering, which shows how noise and disorder can be resources for implementing controlled quantum dynamics of many-body systems. 

\begin{acknowledgments}
We thank T. Northup, V. M. Stojanovi{\'c} and E. Demler for discussions. TF acknowledges support for this work from the Okinawa Institute of Science and Technology Graduate University. HL acknowledges support by a Marie Curie Intra European Fellowship within the 7th European Community Framework Programme. TF and GM acknowledge support from the German Research Foundation (DFG, DACH project ``Quantum crystals of
matter and light") and BMBF (Qu.com).
\end{acknowledgments}




\appendix

\section{Dynamics of the coupled photon-phonon fluctuations}\label{Appendix}

We solve here the linear inhomogeneous system of the Heisenberg-Langevin equations for operators $b_{\alpha}$ and $\delta a$, resulting from Eq. (\ref{fluctuation evolution}); we refer the reader to  \cite{Cormick:2013} for further details. We first introduce dimensionless quadrature operators for field and ions' motion, 
\beq
 X_a = \begin{pmatrix} Q_a\\P_a \end{pmatrix}\,,
 \quad 
 X_\alpha= \begin{pmatrix} Q_\alpha\\P_\alpha \end{pmatrix}\,,
\eeq
with the field quadratures
\beq
\label{eq:field quad}
Q_a = \frac{\delta a + \delta a^\dagger}{\sqrt{2}}, \quad P_a = - {\rm i} \: \frac{\delta a - \delta a^\dagger}{\sqrt{2}}\ ,
\eeq
while $Q_\alpha$ and $P_\alpha$ are defined as
\beq
\label{eq:motion quad}
Q_\alpha = \frac{b_\alpha +b_\alpha^\dagger}{\sqrt{2}}, \quad P_\alpha = - {\rm i} \: \frac{b_\alpha-b_\alpha^\dagger}{\sqrt{2}}\ .
\eeq
We then arrange them together in a column vector,
\beq 
\label{eq:Xdef}
\vec X = \begin{pmatrix} X_a \\ X_1 \\ X_2 \\ \vdots \\ X_N \end{pmatrix} \, ,
\eeq
and write the equations of motion for the fluctuations in the compact form:
\beq
\label{eq:linear compact}
\frac{d\vec X}{dt} = M \vec X + \vec X_{\rm in}(t) \,.
\eeq
Here, $\vec X_{\rm in}$ contains the input noise operators, while
\beq 
\label{eq:M}
M = 
\begin{pmatrix} 
M_a & A_1 & A_2 & \ldots & A_N \\
A_1 & M_1 & 0 & \ldots & 0 \\
A_2 & 0 & M_2 & \dots & 0 \\
\vdots & \vdots & \vdots & \ddots & \vdots \\
A_N & 0 & 0 & \ldots & M_N
\end{pmatrix} \,,                                                 
\eeq
where
\begin{align}
 & M_a = -\kappa \, \mathbb{I} - {\rm i} \, \Delta_{\rm eff} \, \sigma_y\,,\\
 & M_\alpha = -\bar{\Gamma}_\alpha \, \mathbb{I} + {\rm i} \, \omega_\alpha\, \sigma_y\,,\\
 & A_\alpha = -\chi_{\alpha} (\sigma_x - {\rm i} \, \sigma_y) \,,
\end{align}
and $\sigma_{x,y,z}$ are the Pauli matrices. 

For completeness we include in the equations the effect of a noise source, whose occupation number is given by $\bar{N}_{\alpha}$, which  heats the individual vibrational modes with a coupling strength $\bar{\Gamma}_{\alpha}$. However in our calculations we neglect this contribution and set $\bar{\Gamma}_{\alpha}=0$. Thus, the input noise operators correspond only to photonic vacuum fluctuations. Following a standard procedure, we look for the solution of the homogeneous system (\ref{eq:linear compact}) with $\vec X_{\rm in}$ set to zero and then for a particular solution of the inhomogeneous equation. To this aim, one first diagonalizes \eqref{eq:M} and finds the system eigenmodes and eigenvalues, expressed in matrices $B$ and $D$ respectively. 

The solution of this linear system allows one to compute the covariance matrix $\Sigma=\langle X_\delta X_\epsilon + X_\epsilon X_\delta \rangle$, where $\delta$, $\epsilon$ run over all the components of the vector $X$. The steady state of the system is fully described by the covariance matrix evaluated for $t\to\infty$, whose elements are given by:
\begin{equation}
\Sigma_{m n}^{\rm steady} = -2\sum_{\gamma\delta\epsilon}B_{m\gamma}B_{\gamma\delta}^{-1}
\frac{\bar{\Gamma}_{\delta}(2\bar{N}_{\delta}+1)}{D_{\gamma}+D_{\epsilon}}B_{\epsilon\delta}^{-1}B_{n \epsilon}
\label{eq:ent}
\end{equation}
where $\gamma$ and $\epsilon$ run over the collective eigenvectors, while $\delta$ runs over all the components of $X$. The coupling factors for the cavity operators are $\bar\Gamma_a=\kappa$ and we set $\bar{N}_{a}=0$ because we consider only vacuum input noise for the photons. 
The diagonal elements $\langle 2 X_{\alpha}^2 \rangle$ of the covariance matrix determine the final populations of the vibrational modes,
\begin{equation}
n^{\rm steady}_{\alpha}=\frac{1}{2}\left( \frac{\langle 2 Q_{\alpha}^2 \rangle}{2}+\frac{\langle 2 P_{\alpha}^2 \rangle}{2}-1\right)\;,\label{Eqnsteady}
\end{equation}
while the final population of the cavity mode has a similar form.

In order to examine quantitatively the correlations built up between the the cavity and the phonon modes, we study the entanglement between fluctuations, calculated using the logarithmic negativity:
\begin{equation}
E_N=\mbox{max}\{0,-\mbox{ln}(2\nu_-)\}
\label{LNEG}
\end{equation}
where $\nu_{-}$ is the smallest symplectic eigenvalue of the covariance matrix $\tilde\Sigma$ corresponding to the partially-transposed density matrix \cite{Adesso}.

Information about the dynamical properties of the system in the stationary state can be extracted from the spectrum of the field at the cavity output, whose component at frequency $\nu=\omega-\omega_p$ reads $S(\nu)\propto  \langle \tilde{a}_{\rm out}(\nu)^\dagger \tilde{a}_{\rm out} (\nu) \rangle$, with $\tilde{a}_{\rm out}$ the Fourier transform of the field at the cavity output, $a_{\rm out} = a_{\rm in} + \sqrt{2\kappa} \, a$. Using that $a=\bar a+\delta a$, the quantum component of the spectrum reads
\beq
S(\nu) =\frac{\langle \delta \tilde{a} (\nu)^\dagger \delta \tilde{a} (\nu) \rangle}{\bar a^2} \,,
\label{Eq:Snu}
\eeq
where $\delta \tilde{a} (\nu)$ is the Fourier transform of $\delta a$, $\delta \tilde{a} (\nu)^\dagger$ is the Hermitian conjugate of $\delta \tilde{a} (\nu)$, and we omit the Rayleigh peak at $\nu=0$, i.e., $\omega=\omega_p$, which corresponds to the classical part.
Analytical expressions for the spectrum can be found in \cite{Cormick:2013}.

\end{document}